# Listening before Asking: Lived-Experience Advisors as Methodological Partners in Dementia Caregiving Studies

Joy Lai[1][0009-0003-4841-486X], Kelly Beaton[2], David Black[2], and Alex Mihailidis[1][0000-0003-2233-0919]

[1] University of Toronto, Toronto ON, M5S 1A1, Canada
[2] Engagement of People with Lived Experience of Dementia Program/Advisory Group, Canadian Consortium on Neurodegeneration in Aging, Montreal, QC, Canada

`joy.lai@mail.utoronto.ca`

Corresponding author: Joy Lai

**Abstract.** Research with dementia caregivers poses persistent methodological and ethical challenges, particularly when interview-based studies are designed without sufficient grounding in lived caregiving realities. Questions framed through clinical or deficit-oriented assumptions risk alienating participants, undermining rapport, and producing shallow or ethically fraught data. While human-computer interaction (HCI) research increasingly adopts participatory approaches in technology design, participation rarely extends to the design of research methods themselves. This paper examines the role of lived-experience advisors as methodological partners in caregiver interview research. We report on a qualitative study in which two advisors with extensive dementia caregiving experience were engaged prior to fieldwork as methodological partners, extending participatory principles beyond technology design into the design of research methods themselves. Drawing on transcripts of advisor consultations and subsequent interviews with ten caregivers and one person living with dementia, we identify two key methodological contributions of advisor involvement. First, advisors enabled anticipatory validity by surfacing caregiving challenges, ethical sensitivities, and interpretive concerns that later appeared in caregiver interviews, allowing the researcher to enter the field with grounded awareness under constrained recruitment and fieldwork conditions. Second, advisors provided cultural, emotional, and systemic context that improved interpretive sensitivity and helped avoid misreadings. We argue that lived experience functions as methodological infrastructure, extending participatory principles into the design and conduct of research itself, and constituting a generalizable methodological pattern for HCI research with caregivers and other vulnerable or marginalized populations. This model offers a transferable approach to more ethical, rigorous, and contextually grounded HCI research with caregivers and other vulnerable populations.

**Keywords:** Dementia caregiving, lived experience, participatory methods, HCI methodology.

# 1    Introduction

Research with dementia caregivers presents persistent methodological and ethical challenges [1]. Caregiving is emotionally demanding, time-intensive, and deeply contextual, yet interview-based studies often require researchers to enter this space with limited firsthand understanding [2]. As a result, interviews risk "going in blind": asking questions that lack depth, rely on stigmatizing or clinical language, or overlook everyday caregiving practices that participants may not consider significant enough to mention without informed prompting [1, 3]. Even well-intentioned prompts, such as those framed around the "burden of care," can alienate caregivers or disrupt rapport, undermining both data quality and ethical engagement [3]. Because caregivers are often overstretched and recruitment is challenging for multiple reasons, poorly designed interviews can have lasting consequences, deterring future participation and placing additional burden on participants [4].

Within human-computer interaction (HCI), these challenges are increasingly salient [5]. A growing body of work addresses aging, dementia, and caregiving through the design of monitoring systems, assistive technologies, and data-driven decision support [6]. In this context, qualitative accounts are often used to ground, contextualize, or validate AI-driven monitoring and decision-support systems that infer meaning, risk, or "anomalies" from everyday human behavior under conditions of uncertainty and partial data. While HCI has made substantial progress in participatory and co-design methods, these approaches typically focus on the development of technology, not the research methods used to study caregiving itself [7]. Interview guides, analytic frames, and researchers' assumptions are often developed in advance, with little input from those who live the realities under study. This creates a methodological gap: how might participatory principles be extended upstream into the design of research processes themselves?

This paper explores the use of lived-experience advisors, individuals with extended personal histories of dementia caregiving, as methodological partners in interview-based research conducted as part of a broader HCI study on home-based monitoring, anomaly detection, and reminder systems for dementia care [8]. Rather than serving as study participants or co-designers of a technological artifact, advisors in our study were engaged prior to data collection to inform interviewer preparedness, review interview questions, and surface contextual and ethical considerations relevant to this technical context. Our aim was not to replace empirical interviews or evaluate a specific system, but to examine how early and consistent engagement with advisors might improve the quality, integrity, and ethical grounding of interviews conducted within applied, time-constrained research settings.

We report on a qualitative study in which two lived-experience advisors, affiliated with the Engagement of People with Lived Experience of Dementia (EPLED) program, were consulted during the pre-fieldwork phase of a larger project on monitoring and anomaly detection in dementia care [9]. Advisors shared reflections drawn from years of caregiving experience as well as broader insights developed through advocacy and peer engagement. These interactions shaped how the interviewer prepared for fieldwork and engaged participants. Rather than aiming for representativeness or breadth, this engagement prioritized depth of experiential insight and sustained reflexive dialogue, positioning advisors as methodological infrastructure rather than as a sample to be generalized from.

Drawing on advisor consultation transcripts, reflexive memos, and interviews with ten caregivers and one person living with dementia (PLwD), we identify two core methodological contributions of advisor engagement. First, advisors enabled anticipatory validity by surfacing ethical risks, interpretive blind spots, and caregiving priorities before fieldwork began, reshaping interviewer preparedness and reducing the likelihood of misframed or harmful inquiry. Second, advisors provided cultural, emotional, and systemic context that grounded interpretation and supported more accurate, situated analysis. We argue that lived experience functions as methodological infrastructure, extending participatory principles beyond technology design into the core conditions under which qualitative inquiry is conducted.

## 2 Related Work

### 2.1 HCI Research on Dementia and Caregiving

HCI research has increasingly focused on technologies to support PLwD and their caregivers, including monitoring systems, assistive devices, and tools for coordination and decision-making. Prior work has examined caregiver needs around safety, task management, emotional reassurance, and the interpretation of behavioral or environmental data, often emphasizing the delicate balance between technological intervention and human judgment [10]. Studies have also documented the emotional labor and uncertainty caregivers experience when attempting to assess well-being remotely or infer changes in cognitive status from everyday routines [6].

Methodologically, this body of work highlights persistent challenges in studying dementia caregiving. Recruiting caregivers is difficult due to time constraints and emotional demands, and interviews often require substantial contextual grounding to be meaningful [11]. Researchers have noted the ethical risks of deficit-focused or clinical framings, as well as the difficulty of eliciting rich accounts without imposing stigmatizing language or assumptions [3]. While these challenges are well recognized, existing work largely treats them as constraints to be managed rather than as opportunities for methodological innovation.

### 2.2 Participatory and Co-Design Approaches in HCI

Participatory design and co-design are well-established within HCI, particularly in research with older adults, people with disabilities, and marginalized communities [12]. These approaches emphasize collaboration, shared authority, and the inclusion of users' perspectives in shaping technological artifacts. In dementia-related research, participatory methods have been used to co-design interfaces, explore acceptable forms of monitoring, and surface values related to autonomy, privacy, and trust [7].

However, participation in HCI is most often oriented toward technology outcomes rather than research processes [13, 14]. Older adults and caregivers are typically involved as informants, testers, or co-designers of systems, but rarely as partners in shaping interview guides, analytic lenses, or fieldwork practices [12]. As a result, participatory principles often stop at the boundary of the artifact, with core methodological decisions, such as question framing and interpretive assumptions, remaining largely researcher driven.

### 2.3    Lived Experience and Advisory Roles in Research

Outside of HCI, particularly in health, disability, and community-based research, there is a growing recognition of the value of lived experience as a form of expertise [15–17]. Advisory boards, patient partners, and peer consultants are increasingly engaged to guide study design, ensure cultural and ethical appropriateness, and contextualize findings [16, 17]. Such roles are distinct from research participation: advisors contribute higher-level insight drawn from sustained experience and collective knowledge, rather than providing data about individual experiences [18]. While participatory and lived-experience approaches are well established in many qualitative and community-based research traditions, their use as formal methodological input in technical and biomedical HCI research remains comparatively recent and uneven.

Although participatory and lived-experience perspectives are increasingly referenced in HCI research, their integration as methodological partners remains uneven and rarely theorized [12, 14]. Prior work notes that participatory roles are often framed at the level of activities rather than epistemic contribution, with limited documentation of how such involvement shapes research design, analysis, or interpretation [14, 15, 17]. In contrast to advisory roles that primarily focus on governance, consultation, or feedback on predefined study elements, our approach positions lived-experience advisors as epistemic partners who actively shaped methodological assumptions, interpretive framing, and the conditions under which knowledge was produced.

### 2.4    Methodological Gaps and Opportunities

Taken together, prior work establishes both the importance and the difficulty of researching dementia caregiving in HCI [7, 11]. While participatory and lived-experience-informed approaches are valued, they are rarely extended to the design of research methods themselves [14, 18]. This leaves researchers vulnerable to blind spots in language, scope, and interpretation when working with caregivers whose experiences are emotionally complex and structurally constrained [13].

Our work addresses this gap by foregrounding lived-experience advisors as methodological partners in the pre-fieldwork phase of caregiver interview research. Rather than focusing on co-design of a technological artifact, we examine how early engagement with advisors reshaped interviewer preparedness, ethical sensitivity, and interpretive grounding. In doing so, we contribute to ongoing discussions in HCI about participation, reflexivity, and rigor, and offer a concrete, transferable model for integrating lived experience into the design of research processes.

## 3 Methods

### 3.1 Study Overview

This paper reports on a qualitative study examining how lived experience advisors can contribute to the design and conduct of interview-based research with dementia caregivers. The study was part of a broader HCI project exploring technologies for monitoring and anomaly detection in dementia care. Rather than evaluating a technological system, this work focuses on the methodological process through which interviews were planned, conducted, and interpreted, with particular attention to the influence of advisors engaged prior to data collection.

The study unfolded in three phases: (1) pre-fieldwork engagement with lived-experience advisors, (2) semi-structured interviews with caregivers and one PLwD, and (3) thematic analysis of interview transcripts and advisor consultations. While advisor engagement was initially undertaken to support interviewer preparation and ethical sensitivity, this paper examines how its methodological significance became evident through later reflection on the research process and findings.

### 3.2 Lived-Experience Advisors

Two lived-experience advisors were engaged through the EPLED program. Both had extensive experience providing care to a family member with dementia and had participated in peer education and advocacy through EPLED, as well as serving in advisory roles across multiple dementia-related health research studies. Their contributions reflected not only individual caregiving histories but also broader insights shaped by ongoing engagement with caregiver communities.

Advisors were not study participants and did not provide experiential data for thematic analysis. Instead, they were methodological partners who helped surface ethical concerns, identify practical challenges, and guide the refinement of interview tools and framing. Advisors were compensated in line with EPLED guidelines. Given the advisory role focused on sustained reflexive engagement rather than data collection, the number of advisors was determined by depth of collaboration and continuity across the pre-fieldwork phase, rather than by sampling or representativeness.

### 3.3 Advisor Engagement Process

Advisor engagement for the purposes of this methodological analysis occurred prior to caregiver interviews and involved two interrelated activities. Unlike pilot participants or formative interviewees, lived-experience advisors were not engaged to validate interview questions for correctness or completeness, but to surface ethical sensitivities, contextual considerations, and interpretive risks through ongoing dialogue across the pre-fieldwork phase. Meetings were held once a month for over a year, enabling sustained, longitudinal engagement that differentiates this work from short-term participatory or consultative studies and supports the anticipatory validity of advisor input.

**Exploratory Consultations**

Initial meetings were semi-structured and exploratory, focusing on topics such as caregiving routines, uncertainty in monitoring, emotional responses to anomalies, and trust dynamics with personal support workers (PSWs). These conversations were intended to uncover potential blind spots in researcher assumptions and to highlight areas that might require particularly sensitive or careful inquiry.

**Interview Guide Review**

Advisors reviewed a draft interview guide and suggested additional questions and areas of inquiry that had not been previously included. This review also influenced how the guide was used in practice, reinforcing that it should function as a flexible reference rather than a script to be followed verbatim. As a result, the interviewer was better positioned to pursue caregiver-led topics in greater depth, while also refining question phrasing to avoid stigmatizing language, anticipate emotionally charged moments, and use more respectful or culturally appropriate forms of elicitation.

All advisor meetings were transcribed in full, and these transcripts served as the primary source of advisor input. Rather than relying on retrospective notes or post-hoc coding, the researcher treated these transcripts as methodological material, documenting advisor insights in real time, prior to any caregiver interviews being conducted. No formal analysis was applied to advisor transcripts until after the interview phase was completed.

### 3.4    Interview Participants and Recruitment

Following advisor engagement, semi-structured interviews were conducted with ten caregivers and one PLwD. Participants were recruited through Alzheimer Society networks and dementia advocacy groups. Caregivers included spouses, adult children, and grandchildren across a wide age range and varying levels of caregiving intensity.

All interviews were conducted remotely via Zoom, lasted approximately one hour, and were audio-recorded with consent. Interview topics included daily routines, anomaly interpretation, task verification, experiences with care technologies, and emotional responses to uncertainty in care.

### 3.5    Data Analysis

All advisor meetings and caregiver interviews were transcribed verbatim. The data were analyzed using thematic analysis to identify recurring patterns relevant to methodological preparation, caregiver experience, and interpretation. Importantly, advisor transcripts were reviewed for methodological insights only after caregiver interviews had concluded. This sequencing ensured that insights from advisor engagement were not retroactively fitted to caregiver interview data.

Advisor transcripts were analyzed inductively to identify methodological contributions, including warnings, context-setting, and practical insights that shaped interview preparation and interpretation. Caregiver interviews were analyzed to develop themes related to lived experiences of caregiving, anomaly interpretation, and system interaction.

Throughout analysis, the researcher returned to advisor transcripts to assess points of alignment or divergence, recognizing that many of the concerns raised by advisors (e.g., about verification, affect, conversational tone) had been echoed by caregivers in later interviews. These anticipations were not framed as hypotheses, but as grounded insights that had shaped the researcher's pre-fieldwork understanding and later enabled more prepared and sensitive engagement.

### 3.6 Reflexivity and Ethical Considerations

Given the emotional and ethical sensitivity of dementia caregiving, reflexivity was central to this study. Advisor engagement was used proactively rather than reactively, with the intent to identify interactional risks and avoid inadvertently causing distress during interviews. Reflexive decision-making was guided by the goal of minimizing participant burden, avoiding deficit-based language, and increasing interview relevance and comfort.

Importantly, the researcher did not engage in thematic analysis or theory-building during the advisor engagement phase. Advisor input was initially treated as practical preparation and was only later recognized during post-interview analysis as having anticipated many emergent themes. Ethics approval was obtained from the University of Toronto Research Ethics Board, and all participants provided informed consent.

## 4 Findings

### 4.1 Anticipatory Validity

Engaging lived-experience advisors prior to data collection calibrated interviewer preparedness, ethical sensitivity, and interpretive grounding. This anticipatory engagement enabled interviews to focus on uncertainty and emotional labor without requiring participants to correct misframed assumptions or provide foundational explanations during already time-constrained encounters. Unlike familiar practices such as triangulation or member checking, which operate retrospectively on collected data, this form of rigor operates prospectively by reshaping the conditions of inquiry before fieldwork begins.

Advisor engagement prior to data collection contributed to what we describe as anticipatory validity. Rather than generating hypotheses or themes in advance, lived-experience advisors surfaced caregiving concerns, ethical risks, and interpretive frames that later appeared independently in caregiver interviews. These early insights shaped interviewer preparedness and analytic sensitivity, reducing the likelihood of misframed questions or misinterpretation during fieldwork.

Importantly, these anticipations were not treated as findings at the time they were raised. Advisor input was documented in meeting transcripts and memos as practical cautions and contextual grounding. Only during post-interview analysis did the degree of alignment between advisor concerns and caregiver accounts become evident.

**Anticipating Task Verification as a Core Caregiving Challenge**
Before any caregiver interviews were conducted, advisors independently identified task verification as a fundamental and unresolved problem in dementia caregiving. Advisors emphasized that confirming whether activities were actually completed is often impossible without undermining dignity or creating stress for the PLwD. One advisor noted, "But it's a tough thing to really do when the person is there by themselves," and cautioned that verification attempts can easily become inadvertent memory tests.

This framing aligned closely with caregiver interviews. Caregivers consistently described difficulty determining whether tasks had been completed, even when reminders were acknowledged. One caregiver explained, "There would be no definition of done for her. Like she just… wouldn't do anything. She'd say, oh, keys. And then what are keys now?" Others described unreliable self-reporting and reliance on indirect indicators such as environmental cues, including "a wet towel in the hamper" or "a warm microwave."

Advisor input prepared the researcher to treat verification not as a problem of accuracy alone, but as a form of caregiver reassurance under uncertainty. This grounding shaped how interview questions were framed and how caregiver accounts were interpreted, avoiding assumptions that task completion could or should be objectively confirmed. More broadly, this finding aligns with HCI research on uncertainty and human judgment in AI systems, where verification is often an interpretive, human-centered process rather than a purely technical problem of accuracy or ground truth.

**Anticipating Ethical Risks of Language and Question Framing**
Advisors repeatedly warned that question wording and tone provided by reminder systems can provoke anxiety, defensiveness, or confusion, even when questions appear neutral from a research or technical perspective. Advisors emphasized that interrogative phrasing risks stressing PLwD or prompting socially reassuring but inaccurate responses. One advisor explained, "If you say, 'What was the color of the pill,' you're sort of testing their memory."

Caregiver interviews reflected these concerns. Several caregivers stressed that yes/no questions were more manageable than complex prompts, and that overly long or evaluative questions often failed. One caregiver explained, "The only one that I see that is simple enough is, did you finish your breakfast?… By the time I finish the sentence, she might forget the first part." Others contrasted direct questions such as "Did you brush your teeth?" with more indirect alternatives like "Did you use hot or cold water to wet your toothbrush?" which they felt reduced distress.

This anticipatory sensitivity helped the interviewer avoid framing questions or examples in ways that resembled memory testing, relied on evaluative language, or signaled unfamiliarity with dementia care, reducing the risk of derailing interviews into corrective explanation or defensiveness.

**Anticipating Conversational and Engagement Anomalies**

Advisors also highlighted that loss of engagement, flattened responses, or unusual conversational patterns are often more meaningful to caregivers than overt task failures. They cautioned against treating such signals as mood or emotional state, emphasizing that reduced expressiveness or habitual reassurance can mask underlying distress.

Caregiver interviews strongly reflected this framing. Participants described noticing changes in conversational style as early indicators of concern. One caregiver explained, "Have you had your lunch? Yes and no… and he actually picks on the monotonous replies, yes or no," noting that such responses prompted them to spend more time checking in. Another caregiver described tone changes as warning signs: "Flattened speech and anger and all that stuff… sensing agitation, and then it leads to some sort of escalation in her behavior… that would have been helpful to know before it happens."

These accounts aligned with advisor warnings about overinterpreting conversational data. Rather than seeking emotional classification, caregivers valued conversational cues as prompts for attention and reassurance. Advisor input prepared the researcher to interpret these moments not as technical anomalies, but as emotionally and relationally significant signals.

**Anticipating the Need for Summarization Over Raw Data**

Advisors consistently emphasized that caregivers do not benefit from large volumes of raw data. Instead, they highlighted the need for concise, factual summaries that can be shared with clinicians or used to track change over time, framing this as a response to caregiver exhaustion and cognitive overload, as well as the risk of overwhelming clinicians with excessive or poorly curated data.

Caregivers echoed this need. Several described difficulty recalling details when speaking with healthcare providers. One caregiver explained, "Doctors… are asking, how often is she taking her medications? What dose?… I wasn't always the greatest at keeping track of that." This alignment reinforced framing the system as caregiver-facing and interpretive, rather than diagnostic or authoritative.

### 4.2  Contextual Interpretation

In addition to shaping interviewer preparedness, lived-experience advisors contributed to contextual interpretation by providing cultural, systemic, and emotional grounding that informed how caregiver accounts were understood during analysis. Rather than introducing new empirical findings, advisors helped calibrate the interpretive lens through which caregiver narratives were read, ensuring that patterns were situated within caregiving practice rather than treated as technical preferences or isolated anecdotes. This form of contribution was particularly important in areas where meaning depends on tacit knowledge, including time-of-day effects, care system dynamics, language sensitivity, and the emotional burden of uncertainty.

Rather than improving a specific system or interface, advisor engagement shaped how caregiver accounts were interpreted, preventing common misreadings and ensuring that patterns were understood within the emotional, cultural, and systemic realities of caregiving practice. The outcomes of this engagement were therefore methodological rather than evaluative, influencing interpretation, ethical sensitivity, and the boundaries of research claims rather than measurable user experience or design quality. In doing so, this work implicitly critiques standard usability framings in HCI by treating interpretive sensitivity itself as a methodological outcome, rather than as a byproduct of interface evaluation or user feedback.

**Time-of-Day Effects and the Practical Meaning of Sundowning**
Advisors emphasized that caregiving experiences and technology effectiveness vary significantly across the day, particularly in the late afternoon and evening. Drawing on lived experience, they described how confusion, agitation, and disengagement often intensify later in the day, making reminders or prompts less effective and sometimes counterproductive. Advisors framed this not only as a clinical phenomenon, but as a common caregiving reality that shapes when and how interventions can reasonably work.

   Caregiver interviews reflected this same pattern. One caregiver explained, "He gets confused, frustrated, bad mood… then even you remind… [PLwD] may not remember, or listen to it." Others noted that prompts were more effective earlier in the day, when attention and mood were more stable. Advisor input helped the research team interpret these accounts not as scheduling preferences or usability issues, but as reflections of the cognitive and emotional rhythms that can structure dementia care for some PLwD.

**Understanding Trust and Verification in the PSW System**
Advisors also provided essential context about the structure and limitations of PSW systems in their respective provinces, including Ontario and New Brunswick. They explained that while PSWs play a critical role in supporting families, documentation practices and time pressures can create gaps between what is recorded and what actually occurs. Advisors cautioned that families often retain responsibility for verification, even when professional care is present.

   Caregiver interviews echoed this concern. One participant described discovering that recorded tasks did not reflect reality: "She wasn't doing it… she was literally just almost copying and pasting the same notes to save her the time." Advisor explanations of PSW workflows and constraints helped the researcher interpret such accounts not as individual mistrust or dissatisfaction, but as systemic features of fragmented care arrangements. This grounding supported more nuanced analysis of why caregivers emphasized verification, shared records, and caregiver-facing summaries, even in settings with formal support.

**Language Sensitivity and Dignity in Everyday Interaction**
Another area where advisor input shaped interpretation was language sensitivity. Advisors highlighted that seemingly neutral questions can feel blunt, infantilizing, or stressful depending on phrasing, cultural background, and stage of dementia. They emphasized that indirect or personalized questions often preserve dignity and reduce cognitive burden.

This perspective aligned with caregiver accounts. One caregiver contrasted asking "Did you brush your teeth?" with a more specific alternative such as "Did you use hot or cold water to wet your toothbrush?", which they felt was easier to answer without distress. Advisor grounding helped the research team interpret such preferences not as stylistic choices, but as strategies caregivers use to avoid embarrassment, resistance, or confusion. This interpretive lens reinforced the importance of language as an ethical and relational component of caregiving, rather than a purely functional interface detail.

**Interpreting Ambiguity as Emotional Strain, Not Data Absence**
Advisors repeatedly underscored that one of the most difficult aspects of caregiving is not the absence of information, but the emotional strain of uncertainty. They framed monitoring and summaries as sources of reassurance rather than control, emphasizing that caregivers often seek peace of mind rather than a detailed explanation.

Caregivers articulated this anxiety clearly. One participant explained, "They know something's wrong… but they can't see what's going on… the one time I don't pick up the phone, that's gonna be the emergency." Advisor input helped frame such statements not as requests for more data, but as expressions of emotional vigilance and fear of missing a critical moment. This grounding shaped how ambiguity was interpreted during analysis, treating uncertainty as an inherent and emotionally charged feature of caregiving rather than a system failure alone.

**Interpretive Sensitivity as a Methodological Outcome**
Across these examples, lived-experience advisors contributed to what we describe as interpretive sensitivity. By providing cultural references, system-level explanations, and emotional framing in advance, advisors helped the researcher recognize and correctly read nuances in caregiver language, priorities, and concerns.

Without this groundwork, caregiver statements about timing, tone, verification, or trust could have been flattened into technical requirements or usability preferences. Advisor involvement ensured that analysis remained situated in caregiving practice, preserving the relational, emotional, and systemic dimensions that shape how technologies are experienced and evaluated.

## 5    Discussion

This study demonstrates that lived-experience advisors can function as methodological partners in HCI research by shaping how inquiry is prepared, conducted, interpreted, and represented. Rather than contributing empirical data, advisors influenced interviewer preparedness, ethical sensitivity, interpretive grounding, and the framing of research claims. In this section, we discuss how these contributions extend participatory principles into research methods, why anticipatory engagement supports ethical rigor, and how this approach can inform HCI research beyond dementia caregiving.

### 5.1    Lived Experience as Methodological Infrastructure

Our findings show that lived-experience advisors can function as methodological infrastructure rather than as supplementary consultants. Advisor involvement supported anticipatory validity by surfacing caregiving realities, ethical risks, and scope constraints before data collection began. This early grounding helped the researcher recognize and address methodological blind spots that would have been difficult or ethically problematic to identify once interviews were underway. At an institutional level, this model could be operationalized through dedicated advisory budgets, formal recognition of advisory roles in ethics and governance processes, and structured training and compensation frameworks that support sustained engagement across the research lifecycle.

Unlike triangulation or post hoc validation, anticipatory engagement reshaped the conditions of inquiry itself. Advisors did not determine interview themes or analytic outcomes. Instead, they changed how the researcher entered the field by drawing attention to issues such as task verification limits, language sensitivity, and caregiver uncertainty. Treating lived experience as infrastructure highlights its role in supporting rigor and reflexivity, rather than positioning it as an alternative source of empirical data. For HCI researchers working with vulnerable populations, such infrastructure can reduce reliance on inference alone when entering complex lived contexts.

### 5.2    Extending Participatory Design Upstream Into Research Methods

Participatory and co-design approaches in HCI have traditionally focused on shaping technological artifacts. Our findings extend this tradition by showing that research methods themselves are design objects that benefit from participatory engagement. Interview guides, question phrasing, analytic lenses, and claims about impact are all sites where assumptions are embedded and methodological choices shape what knowledge can be produced.

In this study, lived-experience advisors contributed directly to the design of methodological components by shaping how questions were framed, what language was avoided, how ambiguity was interpreted, and how scope and exclusions were articulated. This involvement occurred before interviews, during analysis, and during manuscript development through advisory consultation rather than shared analytic authorship. Rather than treating participation as ending at data collection, this approach demonstrates how lived experience can inform how research is prepared, interpreted, and represented.

### 5.3    Ethical Rigor Through Anticipatory Engagement

A key contribution of advisor involvement was improved ethical rigor, achieved through anticipatory rather than reactive practices. Advisors helped identify interactional risks related to question framing, examples, and tone that could signal unfamiliarity with dementia care, invite evaluative judgments, or derail interviews into correction or defensiveness. Addressing these risks before interviews were conducted reduced the likelihood of avoidable distress and helped preserve the focus and integrity of research encounters.

This anticipatory ethic is especially important in dementia caregiving research, where participants are often overwhelmed, time-constrained, and sensitive to how caregiving practices are represented or evaluated. Advisors' emphasis on tone, simplicity, and restraint illustrates that ethical practice extends beyond consent procedures and institutional review requirements and is enacted moment to moment during interviews. Lived experience played a critical role in surfacing ethical risks that formal protocols alone are unlikely to anticipate.

### 5.4 Research Constraints & Participant Experience: Listening Before Asking

Research with caregivers and other vulnerable populations is often conducted under significant time and recruitment constraints. Studies may have limited enrollment windows, small participant pools, and little opportunity for repeated rounds of data collection. In applied HCI research, interviews are frequently constrained to a single session of limited duration, during which researchers must balance many competing lines of inquiry. Under these conditions, early missteps in framing, language, or assumptions are difficult to correct once an interview is underway.

These constraints are intensified in dementia caregiving research. Recruitment is challenging, and caregivers who choose to participate are often doing so while managing fatigue, stress, and competing responsibilities. Participants are frequently motivated by the hope that research may help reduce uncertainty or workload, creating an ethical obligation to use their time carefully. Ensuring that interviews are respectful, relevant, and well grounded is therefore not only an ethical concern, but a practical one. Poorly framed questions or avoidable confusion risk derailing interviews, shifting time toward correction or explanation, and limiting what can be learned within a single encounter.

Caregiver experiences are also highly heterogeneous. Roles, living arrangements, cultural contexts, and stages of caregiving vary widely, shaping how interviews unfold and what participants prioritize. In this context, interviews cannot reasonably function as practice rounds. Using early participant encounters to surface basic misunderstandings risks losing non repeatable perspectives and placing unnecessary burden on participants whose time and energy are already constrained.

Anticipatory engagement with lived experience advisors helps address these realities by shifting learning upstream. Advisor involvement allowed the researcher to enter interviews with a more accurate and sensitive understanding of caregiving contexts, reducing the need for participants to correct assumptions or provide foundational explanations. This made it possible to use limited interview time more effectively and to focus on interpretation, judgment, and uncertainty rather than remedial clarification.

Listening before asking, in this sense, is a methodological response to real research constraints rather than an aspirational ideal. By investing in anticipatory understanding, researchers can better respect participant contributions, protect scarce interview time, and reduce the risk of avoidable harm in studies where opportunities for engagement may not be repeated.

### 5.5   Interpretive Grounding and the Limits of Automation

Lived-experience advisors also shaped how ambiguity, partiality, and inconsistency were interpreted during analysis. Rather than treating these features as analytic noise or system failure, advisor input helped the researcher interpret them as inherent features of caregiving practice, where uncertainty and judgment are central forms of labor. This grounding was especially important in relation to task verification, conversational anomalies, and caregiver anxiety.

Across advisory meetings, advisors emphasized that caregivers value concise summaries and contextual information over automated interpretation. Caregiver interviews echoed this preference, highlighting the importance of reassurance, verification, and clarity rather than inferred meaning or prescriptive recommendations. Advisor input supported a restrained framing of AI as assistive rather than authoritative, helping the researcher distinguish between useful augmentation and interpretive overreach in contexts where caregiving capacities and constraints shape what automation can reasonably provide.

### 5.6   Participation in Claims-Making and Research Governance

A key implication of this work is that lived-experience advisors shaped not only data collection and interpretation, but also how research claims were bounded and represented. Advisors influenced decisions about what the study could reasonably claim, what should be explicitly excluded, and how feasibility and impact were framed in the manuscript. For example, advisor input prompted explicit avoidance of workload reduction claims and supported framing the study as an exploration of feasibility rather than efficacy.

This involvement positions claims-making as a site of participation rather than a purely researcher-driven activity. By grounding claims in lived caregiving realities, advisor engagement helped the research team avoid overgeneralization and overpromising, particularly in a context where caregivers are already overstretched and sensitive to inflated expectations. Although rarely documented in HCI research, this form of advisory participation functions as a governance mechanism that supports methodological honesty and responsible representation in health and caregiving research.

### 5.7   Implications Beyond Dementia Caregiving

Although this study focuses on dementia caregiving, its methodological implications extend to other HCI domains involving vulnerable or marginalized populations, including accessibility, chronic illness, mental health, and aging-related research. In these contexts, researchers often work across disciplinary boundaries, bringing deep technical expertise into application domains where they may have limited experiential or contextual familiarity. Under time, recruitment, and interpretive constraints, this creates heightened risk of blind spots in framing, stigmatizing language, and overgeneralized claims. These risks are further compounded when researchers assume that personal caregiving experience alone is sufficient to represent broader caregiver perspectives, particularly those of individuals operating outside research and healthcare systems.

Structured advisory partnerships offer a transferable approach for addressing these challenges. Engaging lived-experience advisors early and across the research lifecycle can support interviewer preparedness, ethical sensitivity, interpretive grounding, and accountability, particularly in interdisciplinary research where technical expertise does not necessarily translate into fluency with lived experience. This model does not require large advisory panels. Even a small number of advisors, when engaged intentionally and reflexively, can have substantial methodological impact by shaping how research is prepared, interpreted, and represented.

## 5.8 Limitations and Future Directions

This study involved a small number of advisors and was situated within a specific caregiving and healthcare context. While these constraints limit generalizability, they also clarify the scope of our claims. We argue for the methodological value of lived-experience advisors, not for the universality of specific caregiving insights. Future work could explore advisory partnerships that include PLwD, expand diversity across cultural and socioeconomic contexts, or examine how advisory roles evolve in longitudinal studies.

A further limitation concerns the influence of advisory engagement on the research process itself. Although advisors were not involved in determining interview themes or analytic outcomes, their contributions necessarily shaped interviewer preparedness, attention, and sensitivity. It is therefore possible that advisory input influenced the direction of interviews or interpretation in ways that were not fully conscious or easily disentangled. Rather than treating this as a source of bias to be eliminated, we understand it as an inherent feature of anticipatory engagement, one that trades strict separation for improved ethical sensitivity and contextual grounding. Future work could more explicitly examine how advisory input shapes researcher sensemaking over time, including the limits of reflexive control.

Finally, further research is needed to develop institutional support and recognition for lived-experience advisors, including compensation, authorship, and governance structures. As participatory approaches expand, attention to these infrastructural questions will be essential for ethical and sustainable collaboration.

## 5.9 Conclusion

This study demonstrates how lived-experience advisors can function as methodological partners in HCI research by shaping how inquiry is prepared, conducted, interpreted, and represented. By shifting from ad hoc consultation to structured advisory partnerships, participatory principles are extended beyond technology design into the design of research methods themselves. Advisors contributed not by replacing empirical interviews or generating findings, but by reshaping interviewer preparedness, ethical sensitivity, and interpretive grounding under real constraints on time, recruitment, and participant burden.

Recognizing lived experience as methodological infrastructure clarifies how participatory engagement can improve rigor and accountability without requiring co-authorship or shared analysis. In contexts where researchers work across disciplinary boundaries and engage vulnerable populations, anticipatory advisory engagement offers a practical way to reduce blind spots, protect participant time, and support more careful claims-making. Taken together, this work offers a transferable model for more ethically grounded and contextually situated HCI research with caregivers and other vulnerable populations.

**Acknowledgments.** This work was supported by the Canadian Institutes of Health Research (CIHR), AGE-WELL NCE, EPIC-AT Health Research Training Platform, and VHA Home Healthcare. We thank the caregivers and PLwD who generously shared their time and experiences with the research team.

**Disclosure of Interests.** The authors have no competing interests to declare that are relevant to the content of this article.